\documentclass[entropy,article,accept,moreauthors,pdftex,12pt,a4paper]{mdpi}

\lastpage{\pageref*{LastPage}}
\doinum{10.3390/e17074547}
\pubvolume{17}
\pubyear{2015}
\externaleditor{Academic Editor: Jay Lawrence}
\history{ }
\pdfoutput=1

\usepackage{graphicx}
\graphicspath{{./figures/}}
\usepackage{soul}

\Title{Noiseless Linear Amplifiers in Entanglement-Based Continuous-Variable Quantum Key Distribution}

\Author{Yichen Zhang $^1$, Zhengyu Li $^{2}$, Christian Weedbrook $^{3}$, Kevin Marshall $^4$, Stefano Pirandola $^5$, Song Yu $^{1,}$* and Hong Guo $^{1,2}$}

\address{%
$^{1}$ State Key Laboratory of Information Photonics and Optical Communications, Beijing University of Posts and Telecommunications, Beijing 100876, China \\
$^{2}$ State Key Laboratory of Advanced Optical Communication Systems and Networks, School of Electronics Engineering and Computer Science, Center for Quantum Information Technology and Center for Computational Science and Engineering, Peking University, Beijing 100871, China \\
$^3$ QKD Corp., 60 St. George St., Toronto, M5S 1A7, Canada \\
$^4$ Department of Physics, University of Toronto, Toronto, M5S 1A7, Canada \\
$^5$ Department of Computer Science, University of York, Deramore Lane, York YO10 5GH, UK \vspace{-12pt}}

\corres{E-Mail: yusong@bupt.edu.cn.}

\abstract{We propose a method to improve the performance of two entanglement-based continuous-variable quantum key distribution protocols using noiseless linear amplifiers. The two entanglement-based schemes consist of an entanglement distribution protocol with an untrusted source and an entanglement swapping protocol with an untrusted relay. Simulation results show that the noiseless linear amplifiers can improve the performance of these two protocols, in terms of maximal transmission distances, when we consider small amounts of entanglement, as typical in realistic setups.}


\begin{document}


\section{Introduction}

Quantum key distribution (QKD)~\cite{Gisin_RevModPhys_2002,Scarani_RevModPhys_2009} is the most practical application in the field of quantum information and enables two distant parties, Alice and Bob, to establish a secret key through insecure quantum and classical channels. The continuous-variable version of quantum key distribution (CV-QKD)~\cite{Braunstein_RevModPhys_2005,Xiang-Bin_PhysReport_2007,Weedbrook_RevModPhys_2012}, an alternative to single-photon-based QKD, has attracted much attention in the past few years~\cite{Weedbrook_RevModPhys_2012,Jouguet_nature_2013}, mainly because it does not require single-photon sources or detectors. The Gaussian-modulated CV-QKD protocols based on coherent states~\cite{Grosshans_PhysRevLett_2002,Grosshans_nature_2003,Weedbrook_PhysRevLett_2004} have been experimentally demonstrated~\cite{Lance_PRL_2005,Lodewyck_PhysRevA_2007,Khan_PhysRevA_2013,Jouguet_nature_2013} and have been shown to be secure against arbitrary attacks in the asymptotic~\cite{Renner_PhysRevLett_2009} and finite-size regimes~\cite{Leverrier_PhysRevLett_2013}. Two-way protocols~\cite{Pirandola_NatPhys_2008,SunMZ_IJQI_2012,My_JPhysB_2014,Weedbrook_PhysRevA_2014} and thermal-state protocols~\cite{Weedbrook_PhysRevLett_2010,Usenko_PhysRevA_2010,Weedbrook_PhysRevA_2012} have been also designed.

However, there still exists a gap between the theoretical security analyses and the practical implementations. Such real-life implementations of CV-QKD systems may contain overlooked imperfections, which might not have been accounted for in the theoretical security proofs, and may provide security loopholes. Recently, various attacks have been proposed and closed, such as wavelength attacks~\cite{Xiangchun_PhysRevA_2013_Wavelength,Jingzheng_PhysRevA_2013_Wavelength,Jingzheng_PhysRevA_2014_Wavelength}, calibration attacks~\cite{Jouguetn_PhysRevA_2013} and local oscillator fluctuation attacks~\cite{Xiangchun_PhysRevA_2013_Local}. One approach to overcome device imperfections is by characterizing the whole practical system and to consider all of the existing loopholes. Although some potential loopholes have been discovered and then closed using this approach, it is difficult to find all of the loopholes in practical CV-QKD systems, because the number of loopholes is theoretically infinite.

Another approach is by establishing a full device-independent CV-QKD protocol like its discrete-variable counterpart~\cite{Acin_PhysRevLett_2007}, which is based on the violation of a Bell inequality~\cite{Brunner_arXiv_2013}. Recently, there has been work to build various device-independent CV-QKD protocols, including schemes which are both one-sided~\cite{Walk_arXiv_2014_DICVQKD} and fully device independent~\cite{Christian_arXiv_2014_DICVQKD}. The goal of full device-independent QKD is the removal of the requirement that Alice and Bob need to trust their devices.

In this paper, we consider two kinds of entanglement-based CV-QKD protocols in untrusted scenarios: an entanglement distribution protocol with an untrusted source and an entanglement swapping protocol with an untrusted relay. The latter protocol is inspired by~\cite{Pirandola_PhysRevLett_2012_MDI} and corresponds to the entanglement-based version of the CV-QKD protocols described in~\cite{Zhengyu_PhysRevA_2013,SS-MDI_PhysRevA_2014,Pirandola_arXiv_2013,Ottaviani_PhysRevA_2015}. In particular, we consider a symmetric formulation where the two legitimate partners both modify their data during the classical data post-processing stage.

To improve the maximal transmission distances of these two schemes, we consider the use of two noiseless linear amplifiers (NLAs)~\cite{Xiang_nature_2010}, one at Alice's side and one at Bob's side. We show that the practical example of the CV-QKD protocol with an untrusted source, {\em i.e.}, the entanglement-in-the-middle protocol~\cite{Weedbrook_PhysRevA_2013}, improves by placing two NLAs at the output of the quantum channel at both Alice's and Bob's side. Additionally, the maximal transmission distances of the untrusted relay scheme are also improved using this same method. Previously, a similar method had only been analysed for the case of the one-way CV-QKD protocol~\cite{Blandino_PhysRevA_2012,Jaromir_PhysRevA_2012,Walk_PhysRevA_2012}. These improvements are found in the regime of small entanglement, which is typical in realistic implementations. It is also found that placing only one NLA at the non-reconciliation side (Alice's side for reverse reconciliation and Bob's side for direct reconciliation) has a greater improvement than placing it at the opposite side.

This paper is organized as follows. In Section~\ref{sec:2}, we introduce the two entanglement-based CV-QKD protocols. In Section~\ref{sec:3}, we show that we can improve the performance of these protocols by using NLAs. Our conclusions are drawn in Section~\ref{sec:4}.

\section{\label{sec:2} Entanglement-Based CV-QKD Protocols}

In this section, we begin by describing the two entanglement-based CV-QKD protocols: entanglement-based protocols with an untrusted source and entanglement-based protocol with an untrusted relay, which can also be thought of as entanglement distribution and entanglement swapping protocols, respectively. We then outline the secret key rates for these protocols in the presence of collective Gaussian attacks.

\subsection{Entanglement Distribution: Entanglement-Based Protocols with an Untrusted Source}

\begin{figure}[H]
\center
\includegraphics [height=1.3in]{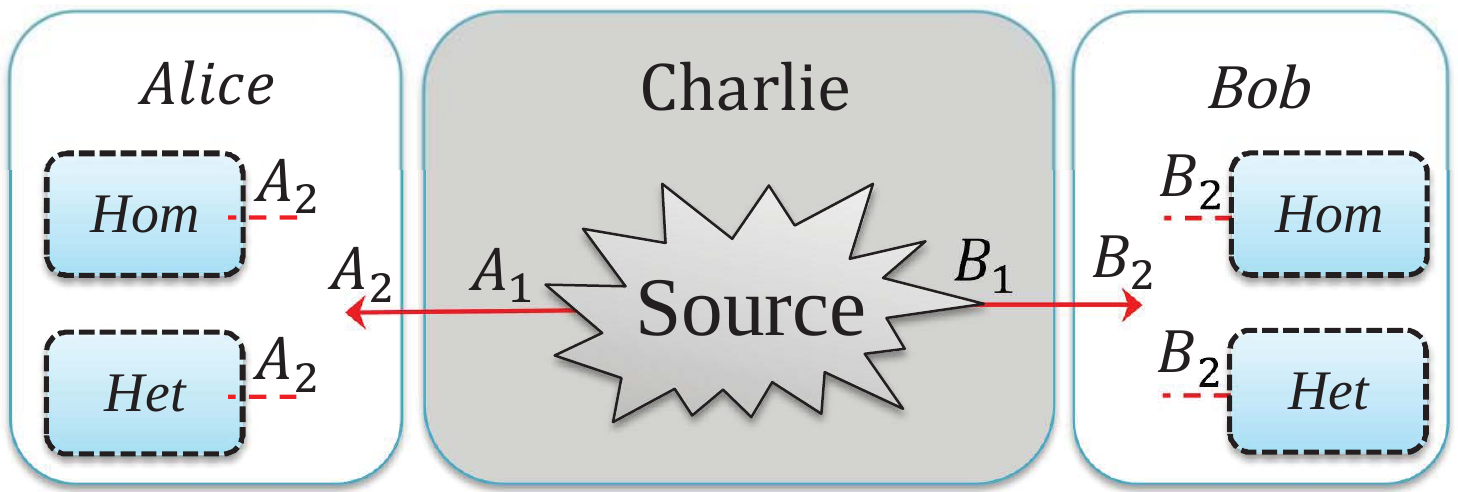}
\caption{Schematic of the continuous-variable version of quantum key distribution (CV-QKD) protocols with an untrusted source. Both the entangled Gaussian source and the quantum channels are fully controlled by Eve. However, Eve has no access to the apparatuses in Alice's and Bob's stations. Alice and Bob can perform either homodyne (Hom) or heterodyne (Het) detection, using either direct or reverse reconciliation.
}\label{fig1}
\end{figure}

The schematic of the entanglement-based CV-QKD protocol with an untrusted source is illustrated in Figure~\ref{fig1} and can be described as follows:

{ Step 1}: The untrusted third party, Charlie, initially prepares an entangled source. He sends one mode $A_1$ to Alice through Channel 1 and sends the other mode $B_1$ to Bob through Channel 2, where Eve may perform her attack.

{ Step 2}: Alice and Bob perform either a homodyne (switching) (Hom) or a heterodyne (no switching) (Het) measurement on the received modes $A_2$ and $B_2$. Once Alice and Bob have collected a sufficiently large set of correlated data, they proceed with classical data post-processing, namely error reconciliation and privacy amplification. The reconciliation can be performed in one of two ways: either direct reconciliation (DR)~\cite{Grosshans_PhysRevLett_2002} or reverse reconciliation (RR)~\cite{Grosshans_nature_2003}.

Since the untrusted Charlie could be completely controlled by the eavesdropper, the original source $\rho_{A_1B_1}^{\left(n\right)}$ (where $n$ denotes the number of quantum signals exchanged during the protocol) is not important to Alice and Bob. What matters is the final state $\rho_{A_2B_2}^{\left(n\right)}$ before their measurements. Here, we assume that the final state $\rho_{A_2B_2}^{\left(n\right)}$ is a `collective source' to simplify the problem, which means Alice and Bob get the same quantum state $\rho_{A_2B_2}$ each time, so that $\rho_{A_2B_2}^{\left(n\right)} = \rho_{A_2B_2}^{ \otimes n}$. The asymptotic secret key rates ${K_{DR}}$ for direct reconciliation and ${K_{RR}}$ for reverse reconciliation are given by~\cite{Devetak_ProcRSoc_2005}:

\begin{equation}\label{e1}
\left\{ \begin{array}{l}
 {K_{DR}} = \beta I\left( {A:B} \right) - \chi \left( {A:E} \right) \\
 {K_{RR}} = \beta I\left( {A:B} \right) - \chi \left( {B:E} \right) \\
 \end{array} \right.,
\end{equation}
where $\beta \in[0,1]$ is the reconciliation efficiency, $I(A:B)$ is the classical mutual information between Alice and Bob, $\chi(A:E)$ and $\chi(B:E)$ are the Holevo quantities~\cite{Nielsen_QCQI}:
\begin{equation}\label{e2}
\left\{ \begin{array}{l}
 \chi \left( {A:E} \right) = S\left( {{\rho _E}} \right) - \sum\nolimits_x {p\left( x \right)S\left( {{\rho _{E|x}}} \right)} \\
 \chi \left( {B:E} \right) = S\left( {{\rho _E}} \right) - \sum\nolimits_y {p\left( y \right)S\left( {{\rho _{E|y}}} \right)} \\
 \end{array} \right.,
\end{equation}
where $S(\rho)$ is the von Neumann entropy of the quantum state $\rho$, $x$ and $y$ are Alice's and Bob's measurement results obtained with probability $p\left( x \right)$ and $p\left( y \right)$, $\rho _{E|x}$ and $\rho _{E|y}$ are the corresponding state of Eve's ancillas and ${\rho _E} = \sum\nolimits_x {p\left( x \right){\rho _{E|x}}}$ and ${\rho _E} = \sum\nolimits_y {p\left( y \right){\rho _{E|y}}}$ are Eve's average states for DR
 and RR
, respectively. Unless both Alice and Bob performed heterodyne measurements, they first apply a sifting process, where they compare the chosen measurement quadrature ($\hat x$ or $\hat p$) and only keep the data for which the quadratures match. Here, we use $x$ and $y$ to represent Alice's and Bob's measurement results, respectively, for both homodyne and heterodyne measurements.

Note that these secret key rates could be modified to take finite-size effects into consideration. For simplicity, here we only consider the asymptotic secret key rates, \textit{i.e}., achieved in the limit of infinite rounds of the protocol. Firstly, Eve is able to purify the whole system $\rho_{A_2 B_2}$ to maximize her information, {\em i.e.}, we have $S\left( {{\rho _E}} \right) = S\left( {{\rho _{{A_2}{B_2}}}} \right)$. Secondly, after Alice's projective measurement resulting in $x$, the system $\rho_{B_2 E}$ is pure, so that ${S\left( {E|x} \right) = S\left( {{B_2}|x} \right)}$ for DR and ${S\left( {E|y} \right) = S\left( {{A_2}|y} \right)}$ for RR. Thus, $\chi \left( {A:E} \right)$ and $\chi \left( {B:E} \right)$ become:
\begin{equation}\label{e3}
\left\{ \begin{array}{l}
 \chi \left( {A:E} \right) = S\left( {{\rho _{{A_2}{B_2}}}} \right) - \sum\nolimits_x {p\left( x \right)S\left( {{\rho _{{B_2}|x}}} \right)} \\
 \chi \left( {B:E} \right) = S\left( {{\rho _{{A_2}{B_2}}}} \right) - \sum\nolimits_y {p\left( y \right)S\left( {{\rho _{{A_2}|y}}} \right)} \\
 \end{array} \right..
\end{equation}

In practical experiments, we calculate the covariance matrix ${{\gamma _{{A_2}{B_2}}}}$ of correlated variables from randomly-chosen samples of measurement data. According to the optimality of collective Gaussian attacks~\cite{Navascu��s_PhysRevLett_2006,Garc��a-Patr��n_PhysRevLett_2006}, we therefore assume that the final state ${{\rho _{{A_2}{B_2}}}}$, shared by Alice and Bob, is Gaussian to minimize the final secret key rates. If the entangled source is Gaussian, one can show that there exists a Gaussian channel mapping the initial state to the final state: this means that there exists a Gaussian attack that is optimal~\cite{Navascu��s_PhysRevLett_2006,Garc��a-Patr ��n_PhysRevLett_2006}. If the entangled source is non-Gaussian, it is an open question whether the optimal attack is Gaussian or not. However, whether Eve's attack is Gaussian or not, we can always bound the information available to Eve by assuming the final state is Gaussian.

Thus, the entropies $S({\rho _{{A_2}{B_2}}})$, $\sum\nolimits_x {p\left( x \right)S\left( {{\rho _{{B_2}|x}}} \right)}$ and $\sum\nolimits_y {p\left( y \right)S\left( {{\rho _{{A_2}|y}}} \right)}$ can be calculated using the covariance matrices ${\gamma _{{A_2}{B_2}}}$ characterizing the state ${\rho _{{A_2}{B_2}}}$, ${{\gamma _{{B_2}|x}}}$ characterizing the state ${{\rho _{{B_2}|x}}}$ and ${{\gamma _{{A_2}|y}}}$ characterizing the state ${{\rho _{{A_2}|y}}}$. The Holevo quantities become:
\begin{equation}\label{e4}
\left\{ \begin{array}{l}
 \chi \left( {A:E} \right) = \sum\limits_{i = 1}^2 {G\left( {\frac{{{\lambda _i} - 1}}{2}} \right) - G\left( {\frac{{{\lambda _3} - 1}}{2}} \right)} \\
 \chi \left( {B:E} \right) = \sum\limits_{i = 1}^2 {G\left( {\frac{{{\lambda _i} - 1}}{2}} \right) - G\left( {\frac{{{\lambda _4} - 1}}{2}} \right)} \\
 \end{array} \right.,
\end{equation}
where $G(x) = (x + 1)\log_2 (x + 1) - x\log_2 x$, ${\lambda _{1, 2}}$ are the symplectic eigenvalues of the covariance matrix ${{\gamma _{{A_2}{B_2}}}}$ and ${\lambda _3}$, ${\lambda _4}$ are the symplectic eigenvalues of the covariance matrices ${{\gamma _{{B_2}|x}}}$ and ${\gamma _{{A_2}|y}}$~\cite{Weedbrook_RevModPhys_2012}.

In particular, a practical example of the CV-QKD protocol with an untrusted source is the entanglement-in-the-middle protocol~\cite{Weedbrook_PhysRevA_2013}, in which the source is assumed to be a two-mode squeezed vacuum state. The latter numerical simulations are also based on this specific example.

\subsection{Entanglement Swapping: Entanglement-Based Protocol with an Untrusted Relay}

\begin{figure}[H]
\center
\includegraphics[width=\textwidth]{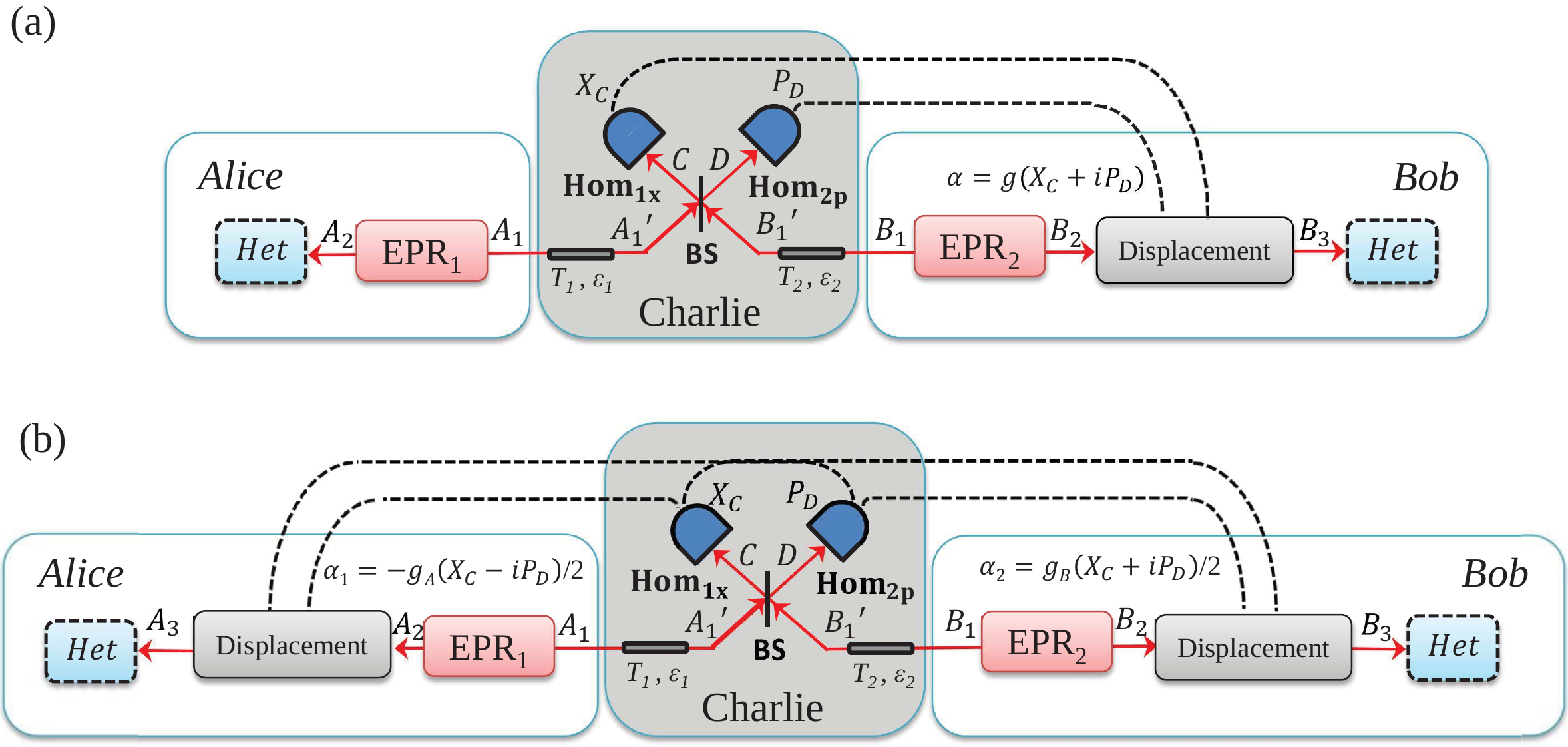}
\caption{(\textbf{a}) Entanglement-based CV-QKD protocol with an untrusted relay where the displacement operator is placed at Bob's side. (\textbf{b}) Entanglement-based scheme where the displacement operator is placed at both Alice's and Bob's sides.
}\label{fig2}
\end{figure}

The schematic of the entanglement-based CV-QKD protocol with an untrusted relay is shown in Figure~\ref{fig2}a. This is inspired by the scheme of~\cite{Pirandola_PhysRevLett_2012_MDI} and represents a modified entanglement-based version of the CV-QKD protocols proposed by~\cite{Zhengyu_PhysRevA_2013,SS-MDI_PhysRevA_2014,Pirandola_arXiv_2013,Ottaviani_PhysRevA_2015}. It can be described as follows:

{Step 1}: Alice and Bob both generate an Einstein--Podolsky--Rosen (EPR), states EPR$_1$ and EPR$_2$, respectively, with variances $V_A$ and $V_B$ and they keep modes $A_2$ and $B_2$ at their respective sides. Then, they send their other modes $A_1$ and $B_1$ to the untrusted third party (Charlie) through two different quantum channels with lengths $L_{AC}$ and $L_{BC}$.

{Step 2}: Charlie combines the received two modes $A'_1$ and $B'_1$ onto a beam splitter (50:50), where we label output modes of the beam splitter as $C$ and $D$. Charlie then measures the \emph{x}-quadrature of mode $C$ and the \emph{p}-quadrature of mode $D$ using homodyne detectors and publicly announces the measurement results $x_{C}, p_{D}$ to Alice and Bob through classical channels. After the measurements of modes $C$ and $D$, the two initially independent modes $A_2$ and $B_2$ get entangled if channel noise is not too strong.

{Step 3}: Bob displaces the mode $B_2$ to $B_3$ by the operation $\hat D\left( \beta \right)$ and gets ${\hat\rho _{{B_3}}} = \hat D\left( \beta \right){\hat\rho _{{B_2}}}{{\hat D}^\dag }\left( \beta \right)$, where $\hat\rho_{B}$ represents the density matrix of mode $B$, $\beta = {g\left( {{X_C} + i{P_D}} \right)} $, $\hat D\left( \beta \right) = {e^{\beta {{\hat a}^\dag } - {\beta ^*}\hat a}}$ (${\hat a}^\dag$ and $\hat a$ are the creation and annihilation operators, respectively), and $g$ represents the gain of the displacement. Then Bob measures the mode $B_3$ to get the final data $\left\{x_B, p_B \right\}$ using heterodyne detection. Alice also measures the mode $A_2$ to get the final data $\left\{x_A, p_A \right\}$, again using heterodyne detection.

{Step 4}: Once Alice and Bob have collected a sufficiently large set of correlated data, they use an authenticated public channel to do parameter estimation from a randomly-chosen sample of final data from $\left\{x_A, p_A\right\}$ and $\left\{x_B, p_B\right\}$. Then, Alice and Bob proceed with classical data post-processing to distil a secret key. The reconciliation can also be done in two ways: either DR or RR.

Note that we can put the displacement operator at each side rather than placing it only at Bob's side, which now makes the protocol symmetric (see Figure~\ref{fig2}b). This symmetry allows the CV-QKD protocol with an untrusted relay to have a similar structure with the entanglement-in-the-middle protocol. In this modified protocol, Alice and Bob displace the modes $A_2$ and $B_2$ by the operators $D\left( \alpha_1 \right)$ and $D\left( \alpha_2 \right)$, resulting in ${\rho _{A_3}} = D\left( \alpha_1 \right){\rho _{{A_2}}}{{D}^\dag }\left( \alpha_1 \right)$ and ${\rho _{B_3}} = D\left( \alpha_2 \right){\rho _{{B_2}}}{{D}^\dag }\left( \alpha_2 \right)$, where $\alpha_1 = {-g_A\left( {{X_C} - i{P_D}} \right)/2 } $, $\alpha_2 = {g_B\left( {{X_C} + i{P_D}} \right)/2 } $, and $g_A$, $g_B$ represents the gain of the displacements at Alice's and Bob's side, respectively.

Note that these protocols can completely defeat side-channel attacks provided that Alice and Bob use quantum memories in their private spaces, which is discussed in detail in~\cite{Pirandola_PhysRevLett_2012_MDI}. From this point of view, this makes the CV-QKD protocol with an untrusted relay more secure. The secret key rate for these protocols against a collective attack is similar to Equation~(\ref{e1}) and can be found in~\cite{Zhengyu_PhysRevA_2013,SS-MDI_PhysRevA_2014} in detail. See~\cite{Pirandola_arXiv_2013} for an unconditional security analysis against the most general coherent attacks.

\section{\label{sec:3}Improvement Using Noiseless Linear Amplifiers}

In this section, we place two noiseless linear amplifiers (NLAs), one at each of Alice's and Bob's side, to improve the performance of the two entanglement-based CV-QKD protocols. We begin by introducing the NLA.

\subsection{Noiseless Linear Amplifier}

For Gaussian states, an NLA can, in principle, probabilistically increase the signal-to-noise ratio by increasing the mean values of the quadratures while keeping their variances at the initial level~\cite{Blandino_PhysRevA_2012,XuBingjie_PhysRevA_2013,Tianyi_PhysLettA_2014,Walk_NJP_2013,Bernu_JPB_2014}. The amplification can be described by an operator $\hat C = {g^{\hat n}}$, where $\hat n$ is the number operator in the Fock basis. Such an operator maps $\left| \alpha \right\rangle$ into $\left| {g\alpha } \right\rangle$ with a success probability $P$, {\em i.e.},
\begin{equation}\label{e5}
\hat C\left( {\left| \alpha \right\rangle \left\langle \alpha \right|} \right) = {P}\left| {g\alpha } \right\rangle \left\langle {g\alpha } \right| + \left( {1 - {P}} \right)\left| 0 \right\rangle \left\langle 0 \right|,
\end{equation}
where $g > 1$ is the gain of the amplifier. Only the situations with successful amplification will be used to distil the final secret keys, while the others are discarded.

In a practical experiment, the covariance matrix before passing through two NLAs takes the form ${\gamma}$, which is used to calculate the final secret key rates (${\gamma _{A_2 B_2}}$ for the entanglement distribution protocols (see Figure~\ref{fig1}) and ${\gamma _{A_3 B_3}}$ for the entanglement swapping protocols (see Figure~\ref{fig2}b)). Typically, ${\gamma}$ can be described by the normal form:
\begin{equation}\label{e6}
{\gamma} = \left[ {\begin{array}{*{20}{c}} {a \cdot I_2} & {c \cdot {\sigma _z}} \\ {c \cdot {\sigma _z}} & {b \cdot I_2} \\ \end{array}} \right],
\end{equation}
where ${{{I}_n}}$ is the $n \times n$ identity matrix, and ${\sigma _z}$ = diag (1, -1).


We then exploit the relationship between the covariance matrix ${\gamma}$ and the density matrix ${{\hat \rho }}$ in the Fock state basis \cite{Jaromir_PhysRevA_2012}. The Husimi Q-function of the two-mode state can be described as:
\begin{equation}\label{e7}
Q\left( {\rm{\textbf{R}}} \right) = \frac{{\sqrt {\det {\Gamma }} }}{{{\pi ^2}}}{e^{ - {{\rm{\textbf{R}}}^T}{\Gamma}{\rm{\textbf{R}}}}},
\end{equation}
where $\textbf{R} = \left( {{{\hat x}_A},{{\hat p}_A},{{\hat x}_B},{{\hat p}_B}} \right)$ and ${\Gamma } = {\left( {{\gamma } + {{I}_n}} \right)^{ - 1}}$. Thus, we can find:
\begin{equation}\label{e8}
{\Gamma } = \left[ {\begin{array}{*{20}{c}} {A \cdot I_2} & {C \cdot {\sigma _z}} \\ {C \cdot {\sigma _z}} & {B \cdot I_2} \\ \end{array}} \right],
\end{equation}
with new parameters A, B and C, after the application of an NLA on each side. In the Fock basis, the Husimi Q-function is a degenerating function of the density matrix elements. Thus, we can establish a relationship between elements of the covariance matrix $\Gamma$ and the elements of the normalized density matrix ${\sigma _{jk,lm}} = {{{\rho _{jk,lm}}} \mathord{\left/ {\vphantom {{{\rho _{jk,lm}}} {{\rho _{00,00}}}}} \right. \kern-\nulldelimiterspace} {{\rho _{00,00}}}}$~\cite{Eisert_AnnPhys_2004}. Then, the matrix $\Gamma$ after the two NLAs becomes:
\begin{equation}\label{e9}
{{\Gamma}_{NLA}} = \left[ {\begin{array}{*{20}{c}}
 {\left( {{g_1^2}\left( {A - \frac{1}{2}} \right) + \frac{1}{2}} \right) \cdot I_2} & {{g_1}{g_2}C \cdot {\sigma _z}} \\
 {{g_1}{g_2}C \cdot {\sigma _z}} & {\left( {{g_2^2}\left( {B - \frac{1}{2}} \right) + \frac{1}{2}} \right) \cdot I_2} \\
\end{array}} \right],
\end{equation}
where $g_1$ and $g_2$ are the gains of the NLAs at Alice's and Bob's sides ($g_1=1$ or $g_2=1$ means there is no NLA). Thus, the covariance matrix ${\gamma'}$ after the NLAs can be obtained by:
\begin{equation}\label{e10}
{{\gamma_{NLA}}} = {\left( {{{\Gamma_{NLA}}}} \right)^{ - 1}} - I_4.
\end{equation}

This covariance matrix is used for the calculation of the final key rates in DR and RR, which are reduced according to the total amplification success probability $P_{\rm{total}} = P_A P_{B|A}$, where $P_A$ is the success probability of Alice's NLA and $P_{B|A}$ is the success probability of Bob's NLA given that Alice's amplification succeeded.
Furthermore, considering the trade-off between the fidelity and the success probability of an NLA, a good estimate of the maximal expected success probability for one NLA is given by~\cite{Caves_PhysRevA_2013}:
\begin{equation}\label{e11}
 P = \frac{{{1}}}{{{g^{2N}}}},
\end{equation}
where $N$ is the average photon number of the input state (ensemble) of the NLA. Such an NLA can amplify an input coherent $\left| \alpha \right\rangle $ to the target output state $\left| g \alpha \right\rangle $ with a relatively high fidelity.

\subsection{Entanglement-Based Protocol with an Untrusted Source}

\begin{figure}[H]
\center
\includegraphics[width=0.7\textwidth]{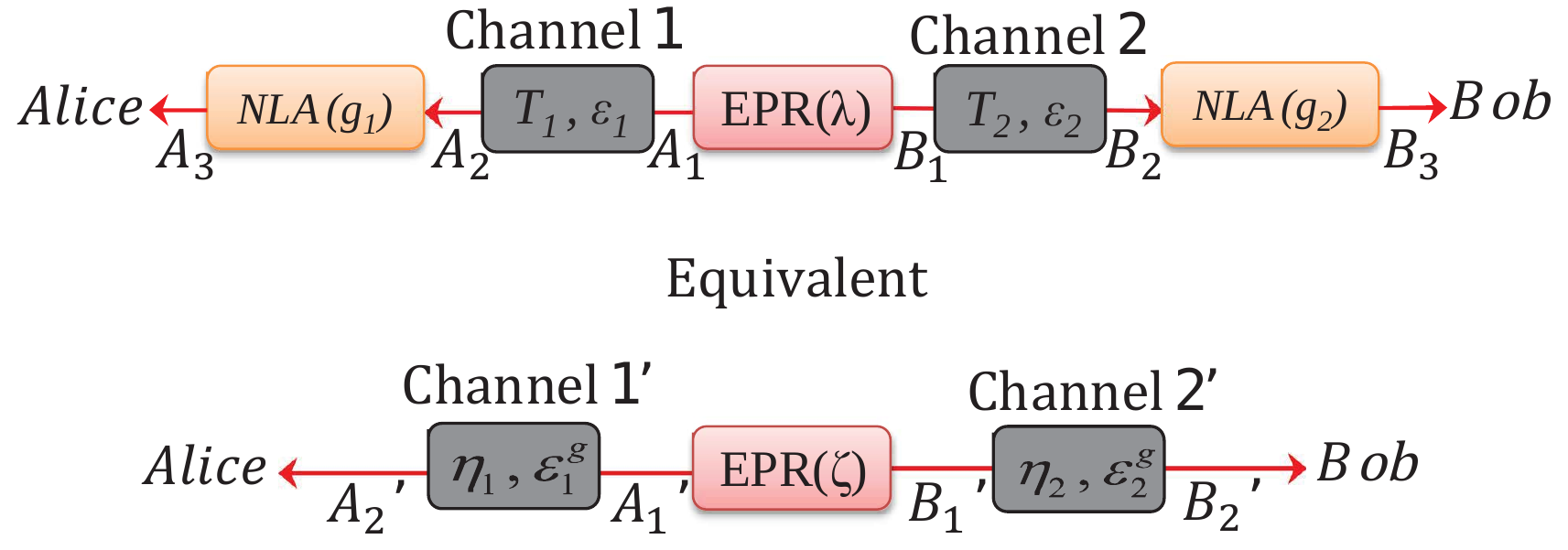}
\caption{Entanglement-in-the-middle protocol. Equivalent channels and squeezing: an Einstein--Podolsky--Rosen (EPR) state $\lambda$ sent through two Gaussian channels of transmittance $T_1$, $T_2$ and excess noise $\varepsilon_1$, $\varepsilon_2$, followed by two successful noiseless linear amplifiers (NLAs), has the same final covariance matrix with a state $\varsigma$ sent through two Gaussian channels of transmittance $\eta_1$, $\eta_2$ and excess noise $\varepsilon_1^g$, $\varepsilon_2^g$, without two NLAs.
}\label{fig3}
\end{figure}

Using the previous method, we can derive the final covariance matrix ${{\gamma}_{A_3 B_3}}$ to calculate the secret key rate. As shown in Figure~\ref{fig3}, we consider a specific example of an entanglement-based protocol with an untrusted source: the EPR in the middle scheme~\cite{Weedbrook_PhysRevA_2013}. This security analysis and latter numerical simulations of this scheme are based on the two independent entangling cloner attacks. This is the most common example of a collective Gaussian attack~\cite{Pirandola_PhysRevLett_2008_Collective}. Alice and Bob both add an NLA before their detectors, which here are assumed to be perfect for simplicity~\cite{Blandino_PhysRevA_2012,XuBingjie_PhysRevA_2013,Tianyi_PhysLettA_2014}.

We can look for equivalent parameters of an EPR state sent through two lossy and noisy Gaussian channels. The covariance matrix ${{\gamma '}_{AB}}\left( {\lambda ,{T_1},{\varepsilon _1},{g_1},{T_2},{\varepsilon _2},{g_2}} \right)$ of the amplified state with an EPR parameter $\lambda$ passing through two channels of transmittance ${T_1},{T_2}$, and excess noise ${\varepsilon _1},{\varepsilon _2}$ followed by two gain efficiencies $g_1,g_2$, is equal to the covariance matrix ${\gamma _{AB}}\left( {\varsigma ,{\eta _1},\varepsilon _1^g,{g_1} = 1,{\eta _2},\varepsilon _2^g,{g_2} = 1} \right)$ of an equivalent system with an EPR parameter $\varsigma$, sent through two channels with parameters $\eta _1$, $\varepsilon _1^g$ and $\eta _2$, $\varepsilon _2^g$, without using NLAs. These parameters are given by:
\begin{equation}\label{e12}
\left\{ \begin{array}{l}
 \varsigma = \lambda \sqrt {\frac{{\left[ {\left( {g_1^2 - 1} \right)\left( {\varepsilon - 2} \right)T - 2} \right] \cdot \left[ {\left( {g_2^2 - 1} \right)\left( {\varepsilon - 2} \right)T - 2} \right]}}{{\left[ {\left( {g_1^2 - 1} \right)\varepsilon T - 2} \right] \cdot \left[ {\left( {g_2^2 - 1} \right)\varepsilon T - 2} \right]}}} \\
 {\eta _1} = \frac{{4Tg_1^2}}{{T\left( {g_1^2 - 1} \right) \cdot \left[ {\left( {g_1^2 - 1} \right)\left( {\varepsilon - 2} \right)\varepsilon T - 4\left( {\varepsilon - 1} \right)} \right] + 4}} \\
 \varepsilon _1^g = \varepsilon - \frac{1}{2}\left( {g_1^2 - 1} \right)\left( {\varepsilon - 2} \right)\varepsilon T \\
 {\eta _2} = \frac{{4Tg_2^2}}{{T\left( {g_2^2 - 1} \right) \cdot \left[ {\left( {g_2^2 - 1} \right)\left( {\varepsilon - 2} \right)\varepsilon T - 4\left( {\varepsilon - 1} \right)} \right] + 4}} \\
 \varepsilon _1^g = \varepsilon - \frac{1}{2}\left( {g_2^2 - 1} \right)\left( {\varepsilon - 2} \right)\varepsilon T \\
 \end{array} \right..
\end{equation}

These could be treated as physical parameters of an equivalent system if they satisfy the following physical constraints:
\begin{equation}\label{e13}
\left\{ \begin{array}{l}
 0 \le \varsigma < 1 \\
 0 \le {\eta _1} \le 1,\varepsilon _1^g \ge 0 \\
 0 \le {\eta _2} \le 1,\varepsilon _2^g \ge 0 \\
 \end{array} \right.
\end{equation}

As shown in Equation~(\ref{e12}), $\lambda$ only affects parameter $\varsigma$, and $\eta _1$, $\varepsilon _1^g$, ${\eta _2}$ and $\varepsilon _2^g$ do not depend on $\lambda$. Thus, the first condition is always satisfied if $\lambda$ is below a limiting value, given by:
\begin{equation}\label{e14}
0 \le \lambda < \sqrt {\frac{{\left( {g_1^2 - 1} \right)\varepsilon T - 2}}{{\left( {g_1^2 - 1} \right)\left( {\varepsilon - 2} \right)T - 2}}} \cdot \sqrt {\frac{{\left( {g_2^2 - 1} \right)\varepsilon T - 2}}{{\left( {g_2^2 - 1} \right)\left( {\varepsilon - 2} \right)T - 2}}} .
\end{equation}

The last two conditions are satisfied if the excess noise $\varepsilon$ is smaller than two and if the gain of the two NLAs is smaller than a maximum value, which depends on the channel parameters $T$ and $\varepsilon$:

\begin{equation}\label{e15}
g_1^{\max } = g_2^{\max } = \sqrt {\frac{{\varepsilon \left[ {T\left( {\varepsilon - 2} \right) + 2} \right] - 2\sqrt {\varepsilon \left[ {T\left( {\varepsilon - 2} \right) + 2} \right]} }}{{T\varepsilon \left( {\varepsilon - 2} \right)}}} .
\end{equation}

Using the previous results, we consider the performance of the CV-QKD protocols with EPR in the middle by placing two NLAs, one at each output of the quantum channels. We calculate the secret key rate $K_{DR}$ as a function of distance $d$ under four situations: without NLAs ($g_1 = 1$, $g_2 = 1$), with only an NLA at Alice's side ($g_2 = 1$), with only an NLA at Bob's side ($g_1 = 1$) and with two NLAs at both sides. The various parameters are chosen from typical experimental values~\cite{Jouguet_nature_2013}: we choose $V = 1.7$, $\beta=0.948$ and $\varepsilon = 0.002$ (where the shot noise variance is normalized to one). The transmittance $T = {10^{ - ad/10}}$, where $a = 0.2$~dB/km is the loss coefficient of the optical fibres and $d$ is the length of the quantum channel.
The total success probability of using two NLAs for the CV-QKD protocols with EPR in the middle is ${P_{\rm{total}}} = {1} / \left({g_1^{2 N_A} g_2^{2 N_{B|A}}} \right)$, where $N_A = T\left(V - 1 + \varepsilon\right) + 1$, $N_{B|A} = T\left(V' -1 + \varepsilon \right) +1 $. Here, $V'$ is the variance of the equivalent EPR when Alice's amplification succeeds, which is given by $V' = {{\left( {1 + {\varsigma ^2}} \right)} \mathord{\left/
 {\vphantom {{\left( {1 + {\varsigma ^2}} \right)} {\left( {1 - {\varsigma ^2}} \right)}}} \right.
 \kern-\nulldelimiterspace} {\left( {1 - {\varsigma ^2}} \right)}}$ provided $g_2 = 1$.

In our analysis, there are eight protocols that depend on Alice's and Bob's measurements (four possibilities) and reconciliation methods (two possibilities, DR or RR). These eight CV-QKD protocols can be divided into four groups whose secret key rate and maximal transmission distance are the same~\cite{Weedbrook_PhysRevA_2013}. When we move the entanglement source into Alice's side, these eight protocols correspond to the entanglement-based version of the eight primary prepare-and-measure CV-QKD protocols, {\em i.e.}, the protocols where Alice and Bob use homodyne detection corresponding to the protocol of~\cite{Cerf_PhysRevA_2001}; the protocols where Alice uses heterodyne detection and Bob uses homodyne detection correspond to the protocol of~\cite{Grosshans_PhysRevLett_2002,Grosshans_nature_2003}; the protocols where Alice uses homodyne detection and Bob uses heterodyne detection correspond to the protocol of~\cite{Patron_PhysRevLett_2009,Pirandola_PhysRevLett_2009_SKeyCapacities}; the protocols where Alice and Bob use heterodyne detection correspond to the protocol of.~\cite{Weedbrook_PhysRevLett_2004}.

\begin{figure}[H]
\center
\includegraphics[width=\textwidth]{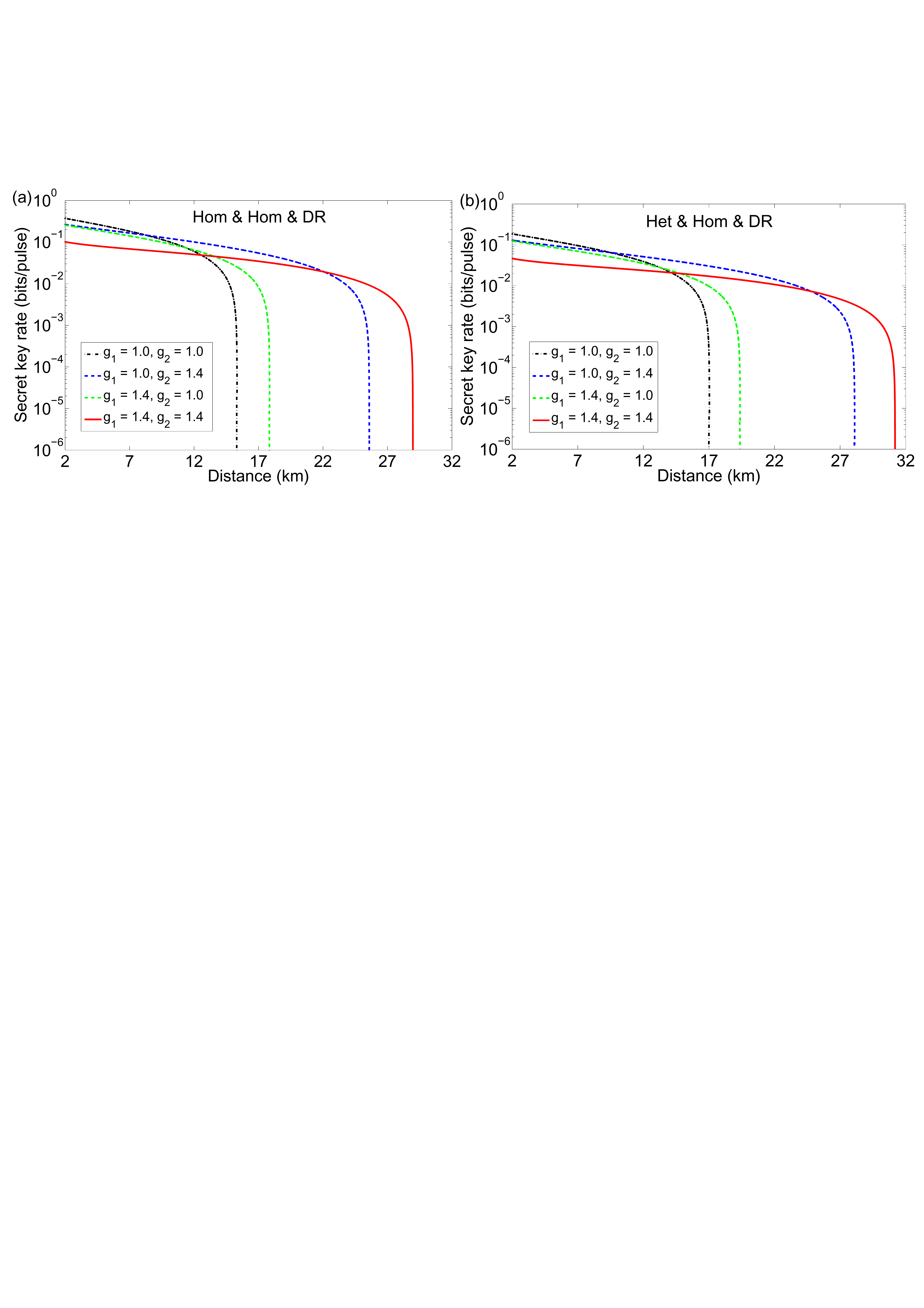}
\caption{Improvement of the CV-QKD protocols with entanglement-in-the-middle. A comparison among the secret key rates for the protocols where (left panel) Alice uses homodyne detection and Bob uses homodyne detection and DR
 (equivalent to Alice using homodyne detection and Bob using homodyne detection and RR
); and (right panel) Alice uses heterodyne detection and Bob uses homodyne detection and DR (equivalent to Alice using homodyne detection and Bob using heterodyne detection and RR), under the following situations: no NLAs ($g_1 = 1$, $g_2 = 1$), using an NLA at Alice's side ($g_2 = 1$), using an NLA at Bob's side ($g_1 = 1$) and using two NLAs at both sides. Here, we use the realistic parameters: $V = 1.7$, $\beta = 0.948$, $\varepsilon = 0.002$ and ${P_{\rm{total}}} = {1} / \left({g_1^{2 N_A} g_2^{2 N_{B|A}}} \right)$.
}\label{fig4}
\end{figure}

\begin{figure}[H]
\center
\includegraphics[width=\textwidth]{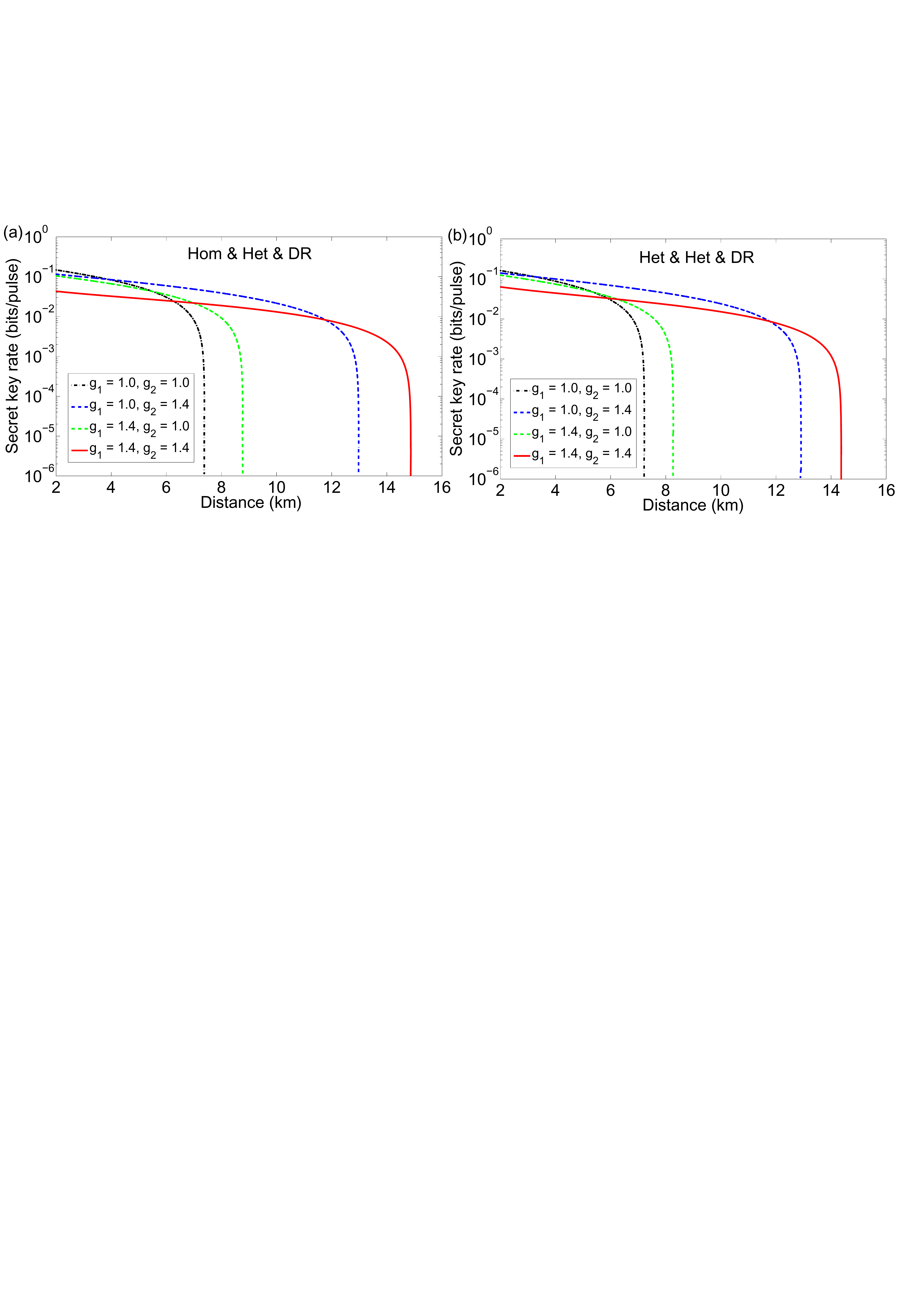}
\caption{Improvement of the CV-QKD protocols with entanglement-in-the-middle. A comparison among the secret key rates for the protocols where (left panel) Alice uses homodyne detection and Bob uses heterodyne detection and DR (equivalent to Alice using heterodyne detection and Bob using homodyne detection and RR); and (right panel) Alice uses heterodyne detection and Bob uses heterodyne detection and DR (equivalent to Alice using heterodyne detection and Bob using heterodyne detection and RR), under the following situations: no NLAs ($g_1 = 1$, $g_2 = 1$), using an NLA at Alice's side ($g_2 = 1$), using an NLA at Bob's side ($g_1 = 1$) and using two NLAs at both sides. Here, we use the realistic parameters: $V = 1.7$, $\beta = 0.948$, $\varepsilon = 0.002$ and ${P_{\rm{total}}} = {1} / \left({g_1^{2 N_A} g_2^{2 N_{B|A}}} \right)$.
}\label{fig5}
\end{figure}

Our simulation results are shown in Figures~\ref{fig4} and \ref{fig5}. We find that the performance of the CV-QKD protocols is improved by placing one NLA at each side and choosing the two gain efficiencies as $g_1 = g_2 = 1.4$. The NLAs enhance the maximal transmission of the protocol, in which Alice is using heterodyne detection and Bob is using homodyne detection with DR, from $17.0$~km to $31.6$~km. Furthermore, we also find that if we only put an NLA at either Alice's or Bob's side, the performance of the protocols can also be improved. For instance, placing an NLA at the non-reconciliation side (Alice's side for RR protocols and Bob's side for DR protocols) has a greater improvement than placing it at the other side. This is because when adding an NLA only at one side (suppose it is on Alice's side), according to Equation (12), the covariance matrix after the application of the NLA has the feature that Alice's equivalent variance is greater than Bob's variance. If considering Alice's part as the reconciliation part, it is similar to the one-way CV-QKD protocol with DR; while, if considering Bob's part as the reconciliation part, it is similar to the one-way CV-QKD protocol with RR. In one-way protocols, the RR protocol usually has a longer transmission distance than the DR protocol. Therefore, in our protocols, placing an NLA at the non-reconciliation side is better than placing it at the reconciliation side. Obviously, the optimal performance of the protocols is achieved by placing two NLAs at each side. However, if we want to reduce the cost and expense and only have one NLA in the deployment, we need to place it at the correct side to have the greatest improvement.
\vspace{-12pt}

Furthermore, as proven in~\cite{Jaromir_PhysRevA_2012,Walk_PhysRevA_2012}, the physical implementation of the NLA could be replaced by a suitable data post-processing (Gaussian post-selection) after the measurement, although provided that certain conditions are met~\cite{Chrzanowski_nature_2014}. Thus, in such cases, we would not need to implement the physical implementation of the NLA, which requires single-photon addition and subtraction, or an auxiliary source of single photons and multiphoton interference~\cite{Jaromir_PhysRevA_2012,Walk_PhysRevA_2012}.

\subsection{Entanglement-Based Protocol with an Untrusted Relay}

\begin{figure}[H]
\center
\includegraphics[width=\textwidth]{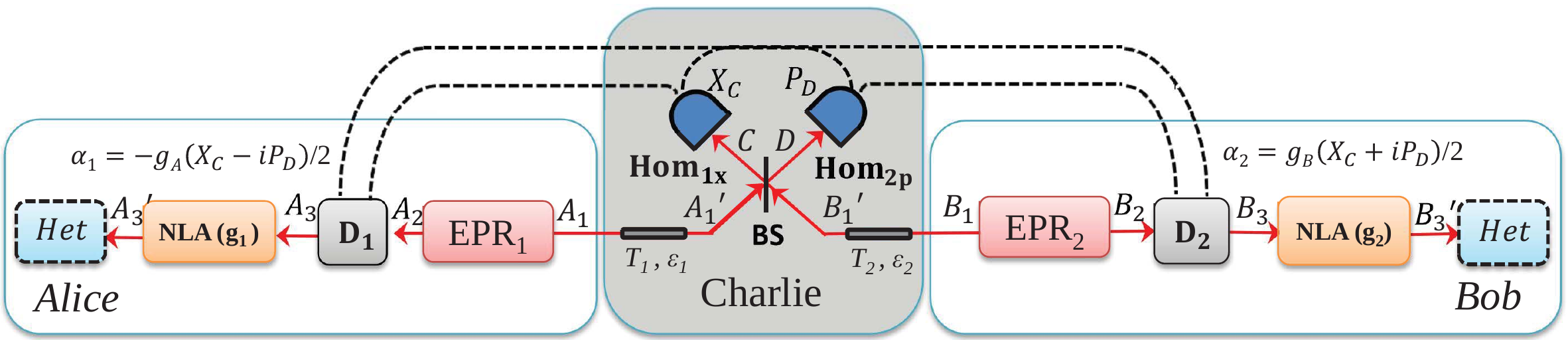}
\caption{Entanglement-based scheme of the modified CV-QKD protocol with an untrusted relay, where a displacement operator $D$ is placed at both Alice's and Bob's sides and the two NLAs are placed before the measurement devices.
}\label{fig6}
\end{figure}

\begin{figure}[H]
\center
\includegraphics[width=\textwidth]{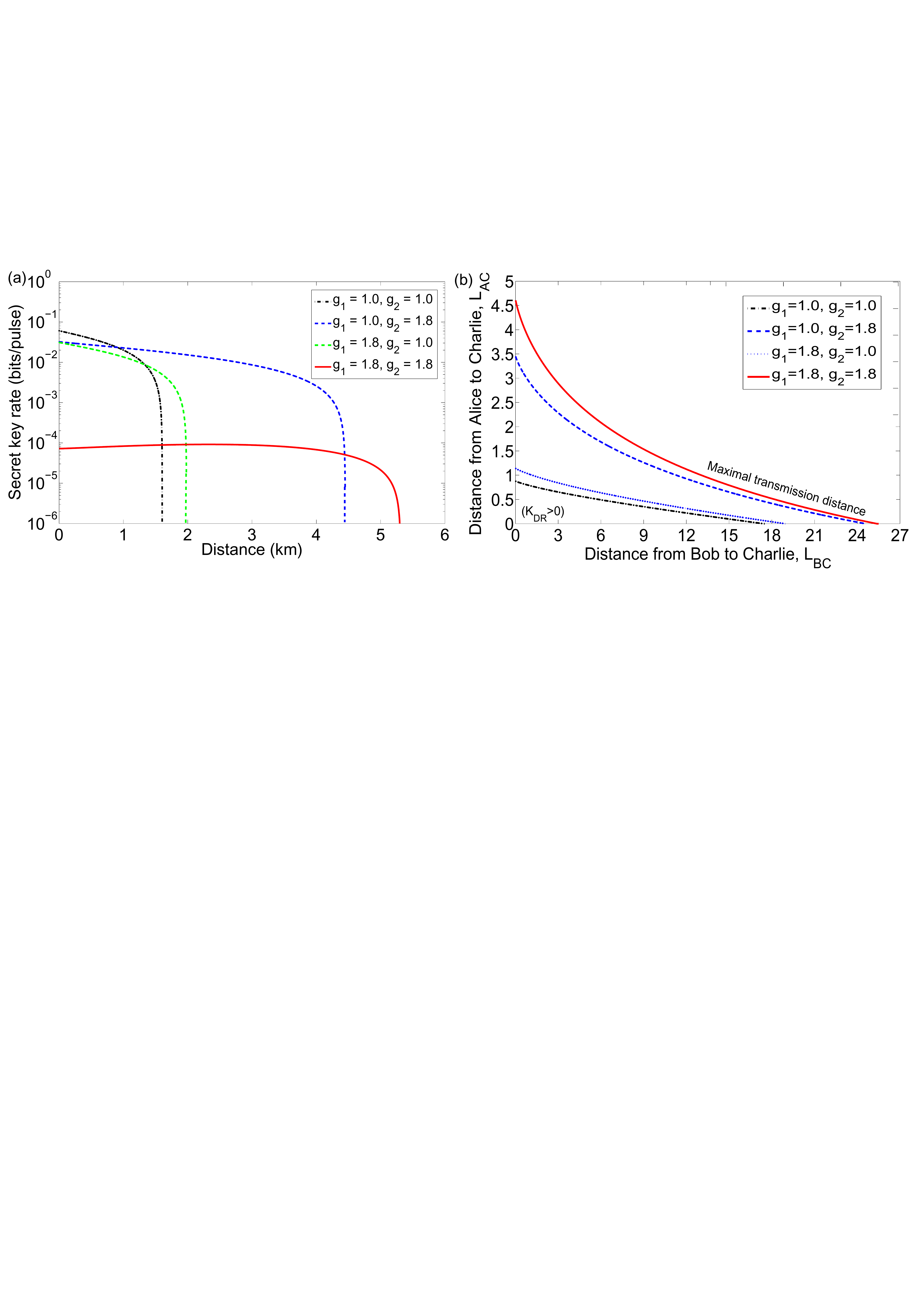}
\caption{Improvement of the modified CV-QKD protocol with an untrusted relay in (\textbf{a})~the symmetric case ({\em i.e.}, $L_{AC} = L_{BC}$) and (\textbf{b}) the asymmetric case ({\em i.e.}, $L_{AC} \neq L_{BC}$). A comparison among the secret key rates in DR under the following situations: no NLAs ($g_1 = 1$, $g_2 = 1$), using an NLA at Alice's side ($g_2 = 1$), using an NLA at Bob's side ($g_1 = 1$) and using two NLAs one at each side. Here, we use the realistic parameters: $V_A = V_B = 1.7$, $\beta = 0.948$, $\varepsilon = 0.002$ and ${P_{\rm{total}}} = {1} / \left({g_1^{2 N_A} g_2^{2 N_{B|A}}} \right)$.
}\label{fig7}
\end{figure}

The improvement seen in the previous section can also be employed in the modified CV-QKD protocol with an untrusted relay. The modified CV-QKD protocol with an untrusted relay is shown in Figure~\ref{fig6} where we place an NLA at both Alice's and Bob's sides. As illustrated in Figure~\ref{fig7}a, the modified entanglement-based protocol can increase the maximal transmission distance when we choose $V = V_A = V_B = 1.7$, $\beta = 0.948$, $\varepsilon = \varepsilon_1 = \varepsilon_2 = 0.002$, ${g_A} = \sqrt {{\rm{ }}{{\left( {{V^2} - 1} \right)} \mathord{\left/ {\vphantom {{\left( {{V^2} - 1} \right)} {\left[ {2{T_1}\left( {V + \varepsilon } \right) + 2\left( {1 - {T_1}} \right)} \right]}}} \right. \kern-\nulldelimiterspace} {\left[ {2{T_1}\left( {V + \varepsilon } \right) + 2\left( {1 - {T_1}} \right)} \right]}}}$, ${g_B} = \sqrt {{\rm{ }}{{\left( {{V^2} - 1} \right)} \mathord{\left/ {\vphantom {{\left( {{V^2} - 1} \right)} {\left[ {2{T_2}\left( {V + \varepsilon } \right) + 2\left( {1 - {T_2}} \right)} \right]}}} \right. \kern-\nulldelimiterspace} {\left[ {2{T_2}\left( {V + \varepsilon } \right) + 2\left( {1 - {T_2}} \right)} \right]}}}$. Under these simulation parameters, the modified entanglement-based protocol in the symmetric case (the distance from Alice to Charlie $L_{AC}$ is equal to the distance from Bob to Charlie $L_{BC}$) can successfully distribute secret keys under such conditions. Then, using the same method as above, we place an NLA at each side to improve its performance; we find an improvement when we set the two gain efficiencies as $g_1 = g_2 = 1.8$. The NLAs enhance the maximal transmission distance of the protocol from $1.6$~km to $5.3$~km in the symmetric case.

Furthermore, for DR, we also find that when Charlie's position is close to Alice, the total maximal transmission distance $L_{AB}$ will increase to a relatively longer distance. Thus, we study the performance of the asymmetric case where $L_{AC} \neq L_{BC}$. As illustrated in Figure~\ref{fig7}b, the total maximal transmission distance increases when $L_{AC}$ decreases. In the asymmetric case, the performance of the modified CV-QKD protocol is also improved by placing two NLAs, one at each side. The maximal total transmission distance of the modified protocol using two NLAs, with gain efficiencies $g_1 = g_2 = 1.8$, is enhanced from $17.5$~km to $25.2$~km in the most asymmetric case ({\em i.e.}, $L_{AC} \approx 0~km$). Here `0~km' indicates that the transmission distance from Alice to Charlie is very short but not exactly zero. In fact, even when Charlie is at Alice's side, there still exists a distance between Alice's laser and the beamsplitter. Therefore, in the numerical simulation although we assume the channel transmittance is $T_1 = 1$, the excess noise $\varepsilon_1$ still exists, and is $\varepsilon_1 = 0.002$.

Note that the sources for Alice and Bob are EPR states. Thus, the protocols can remove side-channel attacks, as discussed in \cite{Pirandola_PhysRevLett_2012_MDI}, which makes the CV-QKD protocol with untrusted relay more secure. Finally, we also find that if we only put an NLA at Alice's or Bob's side, the performance of the protocols can also be improved. This is the same conclusion as before: placing an NLA at the non-reconciliation side (Alice's side for RR protocols and Bob's side for DR protocols) has a greater improvement than placing it at the other side.

\section{\label{sec:4}Conclusion}

In this paper, we have discussed how to improve the performance of two entanglement-based continuous-variable QKD protocols using noiseless linear amplifiers. The first scheme was an entanglement distribution protocol: continuous-variable QKD protocols with an untrusted source, where the entangled source is generated by a third party, but may have actually been created or controlled by the eavesdropper. The second scheme was an entanglement swapping protocol: entanglement-based continuous-variable QKD protocol with an untrusted relay.

By inserting two noiseless linear amplifiers, one at each of Alice's and Bob's side, simulation results show that the proposed method can increase the maximal transmission distances of both protocols in the experimentally-feasible regime of small entanglement, corresponding to small modulation. In fact, in certain situations, we see a doubling of the allowed secure transmission distances. Furthermore, it is also found that placing only one NLA at the non-reconciliation side (Alice's side for reverse reconciliation protocols and Bob's side for direct reconciliation protocols) has a greater improvement than placing it at the other corresponding side.

Future investigations will involve the analysis of the protocols against more general two-mode Gaussian attacks, which are coherent between the two channels connecting the remote parties with the middle source or relay. In fact, as pointed out in~\cite{Pirandola_arXiv_2013}, the unconditional secret-key rate of the relay-based protocol must be derived in the presence of such attacks, which may outperform the collective one-mode Gaussian attacks (based on the use of independent entangling cloners).


\acknowledgments{Acknowledgements}
We thank T. C. Ralph, N.~Walk, A.~Leverrier and M.~Gu for valuable discussions. This work was supported in part by the National Basic Research Program of China (973 Program) under Grants 2012CB315605, in part by the National Science Fund for Distinguished Young Scholars of China (Grant No. 61225003), and in part by the Fund of State Key Laboratory of Information Photonics and Optical Communications. S.~P. would like to thank Engineering and Physical Sciences Research Council (EPSRC) and the Leverhulme Trust for support.

\authorcontributions{Author Contributions}

Yichen Zhang: conception and design of the study, accomplishing formula derivation and numerical simulations, drafting the article.
Zhengyu Li: conception and design of the study, accomplishing formula derivation, checking numerical simulations.
Christian Weedbrook: conception of the study, review of relevant literature, critical revision of the manuscript.
Kevin Marshall: checking formula derivation, critical revision of the manuscript.
Stefano Pirandola: review of relevant literature, critical revision of the manuscript.
Song Yu: review of relevant literature, critical revision of the manuscript.
Hong Guo: review of relevant literature, critical revision of the manuscript.
All authors have read and approved the final manuscript.


\conflictofinterests{Conflicts of Interest}

The authors declare no conflict of interest.

\bibliographystyle{mdpi}
\makeatletter
\renewcommand\@biblabel[1]{#1. }
\makeatother


\begin{thebibliography}{----} 

\bibitem{Gisin_RevModPhys_2002}
Gisin, N.; Ribordy, G.; Tittel, W.; Zbinden, H. Quantum cryptography. {\em Rev. Mod. Phys.} {\bf 2002}, {\em 74}, 145--195.

\bibitem{Scarani_RevModPhys_2009}
Scarani, V.; Bechmann-Pasquinucci, H.; Cerf, N.J.; Du\v{s}ek, M.; L\"utkenhaus, N.; Peev, M. The security of practical quantum key distribution. {\em Rev. Mod. Phys.} {\bf 2009}, {\em 81}, 1301--1350.


\bibitem{Braunstein_RevModPhys_2005}
Braunstein, S.L.; van Loock, P. Quantum information with continuous variables. {\em Rev. Mod. Phys.} {\bf 2005}, {\em 77}, 513--577.


\bibitem{Xiang-Bin_PhysReport_2007}
Wang, X.B.; Hiroshima, T.; Tomita, A.; Hayashi, M. Quantum Information with Gaussian States. {\em Phys. Rep.} {\bf 2007}, {\em 448}, 1--111.


\bibitem{Weedbrook_RevModPhys_2012}
Weedbrook, C.; Pirandola, S.; Garc\'ia-Patr\'on, R.; Cerf, N.J.; Ralph, T.C.; Shapiro, J.H.; Lloyd, S. Gaussian quantum information. {\em Rev. Mod. Phys.} {\bf 2012}, {\em 84}, 621--669.

\bibitem{Jouguet_nature_2013}
Jouguet, P.; Kunz-Jacques,S.; Leverrier, A.; Grangier, P.; Diamanti, E. Experimental demonstration of long-distance continuous-variable quantum key distribution. {\em Nat.~Photon.} {\bf 2013}, {\em 7}, 378--381.



\bibitem{Grosshans_PhysRevLett_2002}
Grosshans, F.; Grangier, P. Continuous variable quantum cryptography using coherent states. {\em Phys. Rev. Lett.} {\bf 2002}, {\em 88}, 057902.

\bibitem{Grosshans_nature_2003}
Grosshans, F.; van Assche, G.; Wenger, J.; Brouri, R.; Cerf, N.J.; Grangier, P. Quantum key distribution using gaussian-modulated coherent states. {\em Nature} {\bf 2003}, {\em 421}, 238-241.

\bibitem{Weedbrook_PhysRevLett_2004}
Weedbrook, C.; Lance, A.M.; Bowen, W.P.; Symul, T.; Ralph, T.C.; Lam, P.K. Quantum cryptography without switching. {\em Phys. Rev. Lett.} {\bf 2004}, {\em 93}, 170504.

\bibitem{Lance_PRL_2005}
Lance, A.M.; Symul, T.; Sharma, V.; Weedbrook, C.; Ralph, T.C.; Lam, P.K. No-switching quantum key distribution using broadband modulated coherent light. {\em Phys. Rev. Lett.} {\bf 2005}, {\em 95}, 180503.

\bibitem{Lodewyck_PhysRevA_2007}
Lodewyck, J.; Bloch, M.; Garc\'{i}a-Patr\'{o}n, R.; Fossier, S.; Karpov, E.; Diamanti, E.; Debuisschert,~T.; Cerf, N.J.; Tualle-Brouri, R.; McLaughlin, S.W.; Grangier, P. Quantum key distribution over $25\phantom{\rule{0.3em}{0ex}}\mathrm{km}$ with an all-fiber continuous-variable system. {\em Phys. Rev. A} {\bf 2007}, {\em 76}, 042305.


\bibitem{Khan_PhysRevA_2013}
Khan, I.; Wittmann, C.; Jain, N.; Killoran, N.; L\"utkenhaus, N.; Marquardt, C.; Leuchs, G. Optimal working points for continuous-variable quantum channels. {\em Phys. Rev. A} {\bf 2013}, {\em 88}, 010302.

%

\bibitem{Renner_PhysRevLett_2009}
Renner, R.; Cirac, J.I. de Finetti representation theorem for infinite-dimensional quantum systems and applications to quantum cryptography. {\em Phys. Rev. Lett.} {\bf 2009}, {\em 102}, 110504.


\bibitem{Leverrier_PhysRevLett_2013}
Leverrier, A.; Garc\'{i}a-Patr\'{o}n, R.; Renner, R.; Cerf, N.J. Security of continuous-variable quantum key distribution against general attacks. {\em Phys. Rev. Lett.} {\bf 2013}, {\em 110}, 030502.


\bibitem{Pirandola_NatPhys_2008}
Pirandola, S.; Mancini, S.; Lloyd, S.; Braunstein, S.L. Continuous-variable quantum cryptography using two-way quantum communication. {\em Nat. Phys.} {\bf 2008} {\em 4}, 726--730.


\bibitem{SunMZ_IJQI_2012}
Sun, M.; Peng, X.; Shen, Y.; Guo, H. Security of a new two-way
continuous-variable quantum key distribution protocol. {\em Int. J. Quantum Inf.} {\bf 2012}, {\em 10}, 1250059.


\bibitem{My_JPhysB_2014}
Zhang, Y.-C.; Li, Z.; Weedbrook, C.; Yu, S.; Gu, W.; Sun, M.; Peng, X.; Guo, H. Improvement of two-way continuous-variable quantum key distribution using optical amplifiers. {\em J.~Phys.~B} {\bf 2014}, {\em 47}, 035501.


\bibitem{Weedbrook_PhysRevA_2014}
Weedbrook, C.; Ottaviani, C.; Pirandola, S. Two-way quantum cryptography at different frequencies. {\em Phys. Rev. A} {\bf 2014}, {\em 89}, 012309.


\bibitem{Weedbrook_PhysRevLett_2010}
Weedbrook, C.; Pirandola, S.; Lloyd, S.; Ralph, T.C. Quantum Cryptography Approaching the Classical Limit. {\em Phys. Rev. Lett.} {\bf 2010}, {\em 105}, 110501.


\bibitem{Usenko_PhysRevA_2010}
Usenko, V.C.; Filip, R. Feasibility of continuous-variable quantum key distribution with noisy coherent states. {\em Phys. Rev. A} {\bf 2010}, {\em 81}, 022318.


\bibitem{Weedbrook_PhysRevA_2012}
Weedbrook, C.; Pirandola, S.; Ralph, T.C. Continuous-Variable Quantum Key Distribution using Thermal States. {\em Phys. Rev. A} {\bf 2012}, {\em 86}, 022318.



\bibitem{Xiangchun_PhysRevA_2013_Wavelength}
Ma, X.-C.; Sun, S.-H.; Jiang, M.-S.; Liang, L.-M. Wavelength attack on practical continuous-variable quantum-key-distribution system with a heterodyne protocol. {\em Phys. Rev. A} {\bf 2013}, {\em 87}, 052309.


\bibitem{Jingzheng_PhysRevA_2013_Wavelength}
Huang, J.-Z.; Weedbrook, C.; Yin, Z.-Q.; Wang, S.; Li, H.-W.; Chen, W.; Guo, G.-C.; Han, Z.-F. Quantum hacking of a continuous-variable quantum-key-distribution system using a wavelength attack. {\em Phys. Rev. A} {\bf 2013}, {\em 87}, 062329.


\bibitem{Jingzheng_PhysRevA_2014_Wavelength}
Huang, J.-Z.; Kunz-Jacques, S.; Jouguet, P.; Weedbrook, C.; Yin, Z.-Q.; Wang, S.; Chen, W.; Guo,~G.-C.; Han, Z.-F. Quantum hacking on quantum key distribution using homodyne detection. {\em Phys. Rev. A} {\bf 2014}, {\em 89}, 032304.


\bibitem{Jouguetn_PhysRevA_2013}
Jouguet, P.; Kunz-Jacques, S.; Diamanti, E. Preventing calibration attacks on the local oscillator in continuous-variable quantum key distribution. {\em Phys. Rev. A} {\bf 2013}, {\em 87}, 062313.


\bibitem{Xiangchun_PhysRevA_2013_Local}
Ma, X.-C.; Sun, S.-H.; Jiang, M.-S.; Liang, L.-M. Local oscillator fluctuation opens a loophole for Eve in practical continuous-variable quantum-key-distribution systems. {\em Phys. Rev. A} {\bf 2013}, {\em 88}, 022339.




\bibitem{Acin_PhysRevLett_2007}
Ac\'{i}n, A.; Brunner, N.; Gisin, N.; Massar, S.; Pironio, S.; Scarani, V. Device-independent security of quantum cryptography against collective attacks. {\em Phys. Rev. Lett.} {\bf 2007}, {\em 98}, 230501.


\bibitem{Brunner_arXiv_2013}
Brunner, N.; Cavalcanti, D.; Pironio, S.; Scarani, V.; Wehner, S. Bell nonlocality. {\em Rev. Mod. Phys.} {\bf 2014}, {\em 86}, 419--478.


\bibitem{Walk_arXiv_2014_DICVQKD}
Walk, N.; Wiseman, H.M.; Ralph, T.C. Continuous variable one-sided device independent quantum key distribution. {\bf 2014}, arXiv:1405.6593.


\bibitem{Christian_arXiv_2014_DICVQKD}
Marshall, K.; Weedbrook, C. Device-independent quantum cryptography for continuous variables. {\em Phys. Rev. A} {\bf 2014}, {\em 90}, 042311.


\bibitem{Pirandola_PhysRevLett_2012_MDI}
Braunstein, S.L.; Pirandola, S. Side-Channel-Free Quantum Key Distribution. {\em Phys. Rev. Lett.} {\bf 2012}, {\em 108}, 130502.


\bibitem{Zhengyu_PhysRevA_2013}
Li, Z.; Zhang, Y.-C.; Xu, F.; Peng, X.; Guo, H. Continuous-variable measurement-device-independent quantum key distribution. {\em Phys. Rev. A} {\bf 2014}, {\em 89}, 052301.


\bibitem{SS-MDI_PhysRevA_2014}
Zhang, Y.-C.; Li, Z.; Yu, S.; Gu, W.; Peng, X.; Guo, H. Continuous-variable measurement-device-independent quantum key distribution using squeezed states. {\em Phys. Rev. A} {\bf 2014}, {\em 90}, 052325.


\bibitem{Pirandola_arXiv_2013}
Pirandola, S.; Ottaviani, C.; Spedalieri, G.; Weedbrook, C.; Braunstein, S.L.; Lloyd, S.; Gehring, T.; Jacobsen, C.S.; Andersen, U.L. High-rate measurement-device-independent quantum cryptography. {\em Nat. Photon.} {\bf 2015}, 397--402.


\bibitem{Ottaviani_PhysRevA_2015}
Ottaviani, C.; Spedalieri, G.; Braunstein, S.L.; Pirandola, S. Continuous-variable quantum cryptography with an untrusted relay: Detailed security analysis of the symmetric conguration. {\em Phys. Rev. A} {\bf 2015}, {\em 91}, 022320.


\bibitem{Xiang_nature_2010}
Xiang, G.Y.; Ralph, T.C.; Lund, A.P.; Walk, N.; Pryde, G.J. Heralded noiseless linear amplification and distilation of entanglement. {\em Nat.~Photon.} {\bf 2010}, {\em 4}, 316--319.


\bibitem{Weedbrook_PhysRevA_2013}
Weedbrook, C. Continuous-variable quantum key distribution with entanglement in the middle. {\em Phys. Rev. A} {\bf 2013}, {\em 87}, 022308.


\bibitem{Blandino_PhysRevA_2012}
Blandino, R.; Leverrier, A.; Barbieri, M.; Etesse, J.; Grangier, P.; Tualle-Brouri, R. Improving the maximum transmission distance of continuous-variable quantum key distribution using a noiseless amplifier. {\em Phys. Rev. A} {\bf 2012}, {\em 86}, 012327.


\bibitem{Jaromir_PhysRevA_2012}
Fiur\'a\v{s}ek, J.; Cerf, N.J. Gaussian postselection and virtual noiseless amplification in continuous-variable quantum key distribution. {\em Phys. Rev. A} {\bf 2012}, {\em 86}, 060302.


\bibitem{Walk_PhysRevA_2012}
Walk, N.; Ralph, T.C.; Symul, T.; Lam, P.K. Security of continuous-variable quantum cryptography with Gaussian postselection. {\em Phys. Rev. A} {\bf 2013}, {\em 87}, 020303.

\bibitem{Devetak_ProcRSoc_2005}
Devetak, I.; Winter, A. Distillation of secret key and entanglement from quantum states. {\em Proc. R. Soc. London Ser. A} {\bf 2005} {\em 461} 207-235.


\bibitem{Nielsen_QCQI}
Nielsen, M.A.; Chuang, I.L. {\em Quantum Computation and Quantum Communication}; Cambridge University Press: Cambridge, UK, 2000.


\bibitem{Navascu��s_PhysRevLett_2006}
Navascu\'{e}s, M.; Grosshans, F.; Ac\'{i}n, A. Optimality of gaussian attacks in continuous-variable quantum cryptography. {\em Phys. Rev. Lett.} {\bf 2006}, {\em 97}, 190502.


\bibitem{Garc��a-Patr��n_PhysRevLett_2006}
Garc\'{i}a-Patr\'{o}n, R.; Cerf, N.J. Unconditional optimality of gaussian attacks against continuous-variable quantum key distribution. {\em Phys. Rev. Lett.} {\bf 2006}, {\em 97}, 190503.







\bibitem{XuBingjie_PhysRevA_2013}
Xu, B.; Tang, C.; Chen, H.; Zhang, W.; Zhu, F. Improving the maximum transmission distance of four-state continuous-variable quantum key distribution by using a noiseless linear amplifier. {\em Phys. Rev. A} {\bf 2013}, {\em 87}, 062311.


\bibitem{Tianyi_PhysLettA_2014}
Wang, T.; Yu, S.; Zhang, Y.-C.; Gu, W.; Guo, H. Improving the maximum transmission distance of continuous-variable quantum key distribution with noisy coherent states using a noiseless amplifier. {\em Phys. Lett. A} {\bf 2014}, {\em 378}, 2808--2812.


\bibitem{Walk_NJP_2013}
Walk, N., Lund, A.P.; Ralph, T.C. Non-deterministic noiseless amplification via non-symplectic phase space transformations. {\em New J. Phys.} {\bf 2013}, {\em 15}, 073014.


\bibitem{Bernu_JPB_2014}
Bernu, J.; Armstrong, S.; Symul, T.; Ralph, T.C.; Lam, P.K. Theoretical analysis of an ideal noiseless linear amplifier for Einstein-Podolsky-Rosen entanglement distilation. {\em J. Phys. B} {\bf 2014}, {\em 47}, 215503.


\bibitem{Eisert_AnnPhys_2004}
Eisert, J.; Browne, D.E.; Scheel, S.; Plenio, M.B. Distillation of continuous-variable entanglement with optical means. {\em Ann. Phys.} {\bf 2004}, {\em 311}, 431--458.


\bibitem{Caves_PhysRevA_2013}
Pandey, S.; Jiang, Z.; Combes, J.; Caves, C.M. Quantum limits on probabilistic amplifiers. {\em Phys. Rev. A} {\bf 2013}, {\em 88}, 033852.


\bibitem{Pirandola_PhysRevLett_2008_Collective}
Pirandola, S.; Braunstein, S.L.; Lloyd, S. Characterization of collective gaussian attacks and security of coherent-state quantum cryptography. {\em Phys. Rev. Lett.} {\bf 2008}, {\em 101}, 200504.


\bibitem{Chrzanowski_nature_2014}
Chrzanowski, H.M.; Walk, N.; Assad, S.M.; Janousek, J.; Hosseini, S.; Ralph, T.C.; Lam, P.K. Measurement-based noiseless linear amplification for quantum communication. {\em Nat.~Photon.} {\bf 2014}, {\em 8}, 333--338.


\bibitem{Cerf_PhysRevA_2001}
Cerf, N.J.; Levy, M.; van Assche, G. Quantum distribution of Gaussian keys using squeezed states. {\em Phys. Rev. A} {\bf 2001}, {\em 63}, 052311.


\bibitem{Patron_PhysRevLett_2009}
Garc\'{i}a-Patr\'{o}n, R.; Cerf, N.J. Continuous-Variable Quantum Key Distribution Protocols Over Noisy Channels. {\em Phys. Rev. Lett.} {\bf 2009}, {\em 102}, 130501.


\bibitem{Pirandola_PhysRevLett_2009_SKeyCapacities}
Pirandola, S.; Garc\'{i}a-Patr\'{o}n, R.; Braunstein, S.L.; Lloyd, S. Direct and Reverse Secret-Key Capacities of a Quantum Channel. {\em Phys. Rev. Lett.} {\bf 2009}, {\em 102}, 050503.






\end{thebibliography}


%


%

\end{document}